\documentclass[twocolumn,showpacs,amssymb,aps]{revtex4-1}

\usepackage{graphicx}
\usepackage{epsfig}
\usepackage{dcolumn}
\usepackage{bm}

\def\lessim{\lower.5ex\hbox{$\; \buildrel < \over \sim \;$}}
\newcommand{\ave}[1]{\left\langle #1 \right\rangle}
\newcommand{\dave}[1]{\ave{\ave{ #1 }}}

\newcommand{\order}[1]{ \mathcal{O} \left( #1 \right) }

\newcommand{\eqcomma}{\phantom{AA},\phantom{AA}}
\newcommand{\volfact}{\frac{4 \pi}{(2 \pi)^3}}
\newcommand{\volfactsq}{\left( \frac{4 \pi}{(2 \pi)^3} \right)^2}

\newcommand{\sfrac}[2]{{\textstyle\frac{#1}{#2}}}
\newcommand\di{\partial}

\begin{document} 
\topmargin -0.8cm\oddsidemargin = -0.7cm\evensidemargin = -0.7cm
\preprint{}

\title{Viscosity of an ideal relativistic quantum fluid: A perturbative study}
\author{Giorgio Torrieri$^a$}
\affiliation{$^a$FIAS, JW Goethe Universitat, Frankfurt }
\date{December 2011}

\begin{abstract}
We show that a quantized ideal fluid will generally exhibit a small but non-zero viscosity due to the backreaction of quantum soundwaves on the background.  We use an effective field theory expansion to estimate this viscosity to leading order in perturbation theory.
We discuss our results, and whether this estimate can be used to obtain a more model-independent estimate of the ``quantum bound'' on the viscosity of physical systems
\end{abstract}

\maketitle
\section{Introduction}
A recent very interesting topic of research has been to transform hydrodynamics, in its Lagrangian formulation, into a quantum field theory \cite{qfluid,qfluid2,qfluideft}.
At first, the whole idea of quantizing ``hot'' hydrodynamics (as opposed to phonons in low-temperature systems: The next section discusses the difference) seems wrong-headed:  ``Everybody knows'' that hydrodynamics is an effective theory describing the infrared limit of a microscopic many-particle system \cite{huang,lifs,kubo1,kubo2,kubo3}.     Such a theory should be inherently classical, since local occupation number is ``large'' and all information about interparticle entanglement is lost in the infrared high temperature limit. 

Nevertheless, several motivations exist for this effort:  The most obvious is ``just to see what happens'':  Lagrangian Hydrodynamics can be rewritten into the form of a field  theory \cite{qfluid2}, so why not quantize it? 

Another motivation is the hope is that this limit might help us understand something about the ``lowest quantum limit for viscosity'' $\eta$ (or, to be more exact, the dimensionless ratio $\eta/s$ where $s$ is the entropy density), widely believed to exist but never conclusively estimated from general principles;  
Existing estimates of this lowest limit are based either on non-rigorous assumptions (for example \cite{daniel}, where the supposedly infinitely strongly coupled fluid is described by a Boltzmann equation) or on theories with classical supergravity holographic limits \cite{4pi,beyond4pi}, well-defined mathematically but whose connection to the real world is problematic.

Moreover, quantizing Lagrangian hydrodynamics could lead to both anomalous and dissipative terms being captured as an effective field theory \cite{qfluideft}, avoiding the ambiguities currently plaguing existing approaches, such as the gradient expansion \cite{jorgegrad}, to derive hydrodynamics as an effective limit of a microscopic theory.

Finally, while the system described here seems removed from anything accessible experimentally, phenomenological repercussions are not a priori excluded: the viscosity of the system created in heavy ion collisions might well be so low \cite{heinzhydro,huovihydro} that a sound wave
\begin{itemize}
\item carrying momentum $p_{sound} \sim (Area) \Delta \rho c_s/k \gg T$ where 
 $\rho,c_s,k,T$ are, respectively, the energy density (and its perturbation $\Delta \rho$), the speed of sound, the wavenumber and the temperature.
  \item Of wavenumber $k \sim p_{sound}$ in natural units
\end{itemize}
will live (not decay to $p_{sound} \sim T$) for a time $\gg 1/T$.
In this regime, {\em some kind of} quantum correction to hydrodynamic evolution becomes mandatory, and quantizing the sound-waves is the simplest ansatz that comes to mind.   

One major stumbling block is that, as \cite{qfluid} has shown, this ``quantum hydrodynamics'' seems at best ambiguous.   Possibly because of turbulence, it is not clear that a stable ground state around which a perturbative expansion can be computed exists in \cite{qfluid}.   Moreover, the approach in \cite{qfluid} and \cite{qfluideft} can not at the moment include dissipative corrections, since by necessity these are non-unitary, and can not be accommodated by the evolution of a pure quantum state.

A possible way to understand these ambiguities was pointed out in \cite{mooresound0,mooresound}:  As is well-known from the non-relativistic hydrodynamics limit, the $\eta \rightarrow 0$ limit is made highly non-trivial due to hydrodynamic fluctuations \cite{lifs}: As $\eta \rightarrow 0$, sound-waves can propagate to asymptotically large distances.  As this limit coincides with the infrared limit in which the transport coefficients are calculated \cite{kubo1,kubo2,kubo3}, these two limits do not in general commute:  In the low viscosity limit the viscosity needs to be {\em renormalized} by a contribution due to sound-waves traveling to asymptotically large distances.    Assuming the maximum wavenumber of sound-waves $k_{max} \sim 1/\eta$, when the ``bare viscosity'' $\rightarrow 0$ the ``renormalized viscosity'' $\rightarrow \eta^{-2}$.    

The contribution of this divergence to the dimensionless parameter $\eta/s$ is vanishing in the limit of large microscopic degeneracy $g$ limit since $ \eta,s \sim g$ while the backreaction terms always $\sim g^0$.   This means backreaction is irrelevant for Yang-Mills type theories with a large number of colors $N_c$, since $g\sim N_c^2$:  $\left. \eta/s \right|_{renormalized} \sim \left. 1/(\eta^2 g)\right|_{bare}$, so the limit depends on ``what diverges faster''.
For instance, in theories with supergravity duals $\left. \eta/s \right|_{renormalized} \sim  1/\left( \order{1} N_c^2 \right) \rightarrow 0$.  The applicability of classical supergravity, therefore, is equivalent to postulating such corrections are negligible.
Physically, the divergence in $g$ can be understood by the requirement that collective degrees of freedom (sound waves etc) carry a negligible amount of entropy w.r.t. microscopic degrees of freedom and hence, do not fluctuate \cite{mooresound0}. 

If $g$ does not diverge (for example, physical QCD at $N_c=3$) and at finite temperature, collective excitations carry a measurable fraction of the entropy w.r.t. microscopic excitations, and hence their fluctuations can not anymore be neglected.
In this regime \cite{mooresound0,mooresound,peralta} show that backreaction is not necessarily negligible. 
An additional problem is that \cite{mooresound,peralta} compares the {\em macroscopic viscosity} with the {\em microscopic} entropy.
To estimate a {\em fully macroscopic} $\eta/s$ with the approach of \cite{mooresound}, one would need to 
\begin{itemize}
\item Estimate the contribution to the {\em entropy density} resulting from an equilibrated ``gas of soundwaves''
\item Take the sound-sound interactions fully into account while calculating viscosity.
\end{itemize}
In this work, address both the open questions in  \cite{qfluid,qfluid2,qfluideft} and in \cite{mooresound} by calculating the ``quantum viscosity'' of an {\em ideal} fluid such as the one of \cite{qfluid,qfluid2,qfluideft}.  There is a hope to do this self-consistently since quantization will allow us to unambiguously assign an entropy and a viscosity to a gas of Lorentz-scalar sound-waves.

That an entropy can be unambiguously assigned after quantization is clear from considering the finite occupation number of each sound mode.   That a viscosity can also be assigned is also clear from considering the Kolmogorov cascade \cite{lifs} evolution of a turbulent system: a classical Kolmogorov cascade, which typically converts higher amplitude lower frequency perturbations into lower amplitude, higher frequency ones, can go on indefinitely since there is no automatic relationship between the wavenumber of the perturbation and its energy.  Quantization will automatically cut the cascade off by assigning more energy to higher frequency modes.   Since in classical hydrodynamics the Kolmogorov cascade is cut-off by viscous effects \cite{lifs}, this quantum cut-off can be understood as an effective quantum viscosity.
   
The full hydrodynamic Lagrangian will also allow us to calculate the interactions between these sound waves by usual Feynman diagram techniques \cite{peskin}, and to see whether quantum corrections tame the divergence pointed out in \cite{mooresound}, in a way that gives a finite renormalized $\eta/s$ (even with $\eta,s$ diverging separately).

Since the theory in \cite{qfluid,qfluid2,qfluideft} is non-renormalizeable, it is not surprising that quantum perturbations give an inherently non-unitary dissipative correction (A similar situation exists in EFTs such as chiral perturbation theory \cite{danielf2,danielf}, where unitarity is restored by hand at each order.   Unlike this theory, however, viscous hydrodynamics is explicitly dissipative, and hence non-unitary).  We nevertheless argue that looking for fixed points in $\eta/s$ when the ``microscopic'' parameters diverge could be used to extract an estimate for the lower quantum limit of $\eta/s$ and its dependence on the equation of state.
\section{The theory}
We consider an ideal quantum fluid of the type described in \cite{qfluid,qfluid2,qfluideft}, and use the notation of these works henceforward.   First, to avoid confusion, we emphasize that this theory is fundamentally different from the usual ``sound quantization'' (``phonons'') associated, for example, with liquid helium and superconductors.
 The latter is usually a system at high chemical potential (high particle density) and very low temperature, so $T/\mu \ll 1$  and the system is essentially a quantum many-body wavefunction.   
Here, there is {\em no} chemical potential since there are no conserved quantum numbers, and all energy is encoded into temperature.   The ``conserved quantity'' that moves around during the fluid's evolution is {\em not} particle number but {\em microscopic entropy} which is non-zero ( as it usually is in a coherent quantum system) because the microscopic degrees of freedom are perfectly thermalized and fully incoherent, an assumption equivalent to that of perfect fluidity.  This work rests on the assumption that {\em macroscopic} degrees of freedom can be meaningfully quantized in this limit.   The ``yes'' answer, assumed here and in \cite{qfluid,qfluid2,qfluideft}, is reasonable simply because, as argued in the introduction, for arbitrarily low viscosities quantum corrections to sound-waves {\em must} make an appearance.  Nevertheless, this assumption has not to our knowledge explored from the ``microscopic QFT'' point of view.

The degrees of freedom of such a chargeless ideal fluid, in the Lagrangian formulation, are the spatial coordinates $\phi_I$ of a fluid volume element comoving with that element.
It can be shown  \cite{qfluid,qfluid2,qfluideft} that, if the Lagrangian is a function of $B$ {\em only}
\begin{equation}
\label{lfluid}
L=F(B) = F \left( \mathrm{det}\left( B_{IJ} \right)\right)
\end{equation}
where
\[\  B_{IJ} =  \partial^\mu \phi_I \partial_\mu \phi_J   \]
the energy momentum tensor will have the form of a relativistic ideal fluid \cite{lifs,weinberg}
\begin{equation}
\label{entensor}
T_{\mu \nu} = (p+\rho) u^\mu u^\nu -p g^{\mu \nu} 
\end{equation}
where the energy density and pressure are
\begin{equation}
\rho = -F(B)\end{equation}
\begin{equation}
 p=F(B)-2 \frac{dF}{dB}B
\end{equation}
and the flow is
\begin{equation}
u^\mu = \frac{1}{6 \sqrt{B}} \epsilon^{\mu \alpha \beta \gamma}\epsilon_{I J K} \partial_\alpha \phi^I  \partial_\beta \phi^J   \partial_\gamma \phi^K 
\end{equation}
The relation  $\rho = -L$ shows that, other than the somewhat unorthodox notation, the approach used here coincides with the usual Lagrangian hydrodynamics used in numerical simulations \cite{lifs,kodama}.

The formulae above can be used to show that 
$\partial_\mu \left( \sqrt{B} u^\mu \right)=0$.  $\sqrt{B}$ is therefore, a locally conserved quantity, the only one of this theory.    We will identify it with the microscpic entropy, up to a ``degeneracy'' constant $g$.
\begin{equation}
\label{sdef}
s = g\sqrt{B}
\end{equation}
Note that $g$ could very well diverge if the ``microscopic'' and ``macroscopic'' degrees of freedom are well-separated (as it does, for example, in theories with a classical gravity dual such as \cite{4pi,beyond4pi}).  We introduce this constant in light of the arguments made in the introductions about balancing microscopic and macroscopic entropy.   As we shall see these issues do {\em not} go away once the fluid is quantized.

The ``equilibrium'' between microscopic  and macroscopic degrees of freedom can then be ensured by the  Gibbs-Duhem relation \cite{huang}, relating entropy $s$,temperature $T$ and enthalpy $p+\rho=w$.
\begin{equation}
\label{gibbs}
s = \frac{dP}{dT} = \frac{p+\rho}{T} = \frac{w}{T} \sim \sqrt{B} 
\end{equation}
the choice of $g$ and the Gibbs-Duhem relation specifies the microscopic temperature as
\begin{equation}
T=\frac{\sqrt{B}(dF/dB)}{g}
\end{equation}
(Note that if energy stays finite but $g$ diverges, $T \rightarrow 0$).

Since everything is formulated via the Lagrangian notation, at this point nothing prevents us from quantizing using the usual sum-over-path prescription and calculating any expectation value of a fully-quantum operator \cite{peskin}.   For example, the Quantum two-point function for the energy-momentum tensor Eq. \ref{entensor} will be
\begin{equation}
\label{ttcorr}
\ave{T_{\mu \nu} (x) T_{\mu \nu} (x')} =\phantom{AAAAAAAAAAAAAAAA}
\end{equation}
\[\ \int \mathcal{D} \phi_{I,J,K} \left(T_{\mu \nu} (x) T_{\mu \nu} (x')\right)_{\phi_{I,J,K}} \exp \left[ - i F \left( B_{\phi_{I,J,K}} \right) \right]  \]
Calculating anything is however a very non-trivial affair, since even the simplest ``ideal gas'' equation of state
\begin{equation}
F(B) = - B^{2/3}
\end{equation}
gives rise to a deeply intractable theory where, as will become shortly clear, the existence of a useful decomposition into free states (an S-matrix definition) is laden with problems.    Perturbatively, however, we can make progress by expanding around a hydrostatic solution where the background is fixed at a constant $\vec{X}_I$ plus transverse and longitudinal ``phonons'' (note the difference between these and what are usually called phonons, explained at the start of this section)
\begin{equation}
\phi_I = \vec{X}_I + \vec{\pi}_L + \vec{\pi}_T
\end{equation}
The symmetries evident in the Lagrangian description make it simple to show that the only ``non-trivial'' excitations are transversely polarized ``vortices'' and longitudinally polarized ``sound-waves''.

A short calculation is however enough to confirm the result, known to anyone familiar with fluids,  that classical vortices in ideal fluids {\em live forever}, carry an arbitrarily small amount of energy and an arbitrarily large amount of momentum, but {\em do not propagate} \cite{lifs} .  Basic quantum field theory, therefore, shows that any scattering approximation breaks down in the presence of vortices, as ``virtual vortices'' can be generated for arbitrarily long times with no energy cost \cite{peskin}.

A somewhat ad hoc but consistent fix to this   \cite{qfluid,qfluid2,qfluideft} is to add a small propagation velocity for vortices by considering the Lagrangian not of a fluid but of a ``soft Jelly''
\begin{equation}
\label{lfluidct}
L_{jelly} = F(B) + \frac{1}{2} c_T^2 \frac{dF}{dB} \sum_I B_{II}
\end{equation}
A fluid can then be thought of as $\lim_{c_T \rightarrow 0} L_{jelly}$.
We note that $\sqrt{B}$ is not anymore a conserved quantity even at the classical level, as we have
\begin{equation}
T_{\mu \nu} \rightarrow T_{\mu \nu}^{ideal} +  c_T^2 B_{II} \left( J_1 u_\mu u_\nu  + J_2 g_{\mu \nu}  \right)
\end{equation}
$ J_1 = \frac{dF}{dB} + B  \frac{d^2 F}{dB^2} \eqcomma J_2 = - \frac{3}{2}\frac{dF}{dB}  $
This is equivalent to shifting $\rho \rightarrow \rho+c_T^2 B_{II} \left( J_1 + J_2 \right), p \rightarrow p - c_T^2 B_{II} J_2$.    The conserved charge corresponding to entropy will correspondingly get a contribution from transverse stresses, 
\begin{equation}
\label{eqsct}
 \delta s \sim c_T^2 g \order{  \frac{B_{II}}{\sqrt{B}}} \left( 1 +  \frac{(d^2 F/dB^2)B}{(dF/dB) } \right)   \end{equation}
This looks somewhat similar to the familiar bulk viscosity term;  However, it is radically different as the equations of $B$ and $B_{II}$ are still non-dissipative.  Eq. \ref{eqsct} will be augmented by additional classical equations of motion for $B_{II}$ derivable from Eq. \ref{lfluidct} via the usual Lagrangian prescription.

Hence, classically, if we adjust the initial conditions to ensure that $ c_T^2 g B_{II}/\sqrt{B} \ll 1$, entropy as defined in Eq. \ref{sdef} will tend to a conserved quantity.   Quantum fluctuations induced by the new term will however remain.

For a fluid with enthalpy $w_0,$ and $B=B_0$, the linearized Lagrangian, up to 4th order, is then 
\begin{equation}
L= w_0 \Big\{ \sfrac12 \dot {\vec \pi}^2 - \sfrac12 c_s^2 [\di \pi]^2  - \sfrac12 c_T^2 [\di \pi^T \di \pi ]
\end{equation}
\[\ + \, \sfrac12 c_s^2 [\di \pi] [\di \pi^2] - \sfrac16 \big( 3 c_s^2 + f_3 \big) [\di \pi]^3 +    \sfrac12(1+c_s^2) \,  [\di \pi] \dot {\vec \pi}^2  - \dot {\vec \pi} \cdot \di \pi \cdot \dot {\vec \pi} \]
\[\ - \, c_s^2 [\di \pi] \det \di \pi   - \sfrac18 c_s^2   [\di \pi^2]^2   + \sfrac14  \big( c_s^2 + f_3 \big) [\di \pi^2][\di \pi]^2  \]
\[\
-\sfrac1{24} \big( 3 c_s^2 +  6 f_3 +f_4 \big) [\di \pi]^4  \]
  \[\ + \, \dot {\vec \pi} \cdot \di \pi^2 \cdot \dot {\vec \pi}  
- (1+c_s^2) [\di \pi] \, \dot {\vec \pi} \cdot \di \pi \cdot \dot {\vec \pi} + \sfrac12 | \di \pi^T \cdot \dot {\vec \pi}|^2 \]
 \[\ \left. +    \,  \sfrac14 \big( (1+3 c_s^2 + f_3) \, [\di \pi]^2 - (1+c_s^2) \,  [\di \pi^2] \big) \dot {\vec \pi}^2  + \sfrac18 (1- c_s^2) \, \dot {\vec \pi}^4 
\right\} \]
(the first line covers the ``free'' kinetic energy terms).
All parameters of the Lagrangian can be derived from $F(B_0)$, the energy density of the hydrostatic background.
\begin{equation}
w_0 = 2 \sqrt{B_0} \left. (dF/dB)  \right|_{B=B_0}
\end{equation}
\begin{equation}
c_s^2 = \left. \frac{2 (d^2 F/dB^2)B+dF/dB}{dF/dB} \right|_{B=B_0}
\end{equation}
\begin{equation}
f_n = \left. \frac{(d^n F/dB^n) B^{n-1}}{dF/dB} \right|_{B=B_0}
\end{equation}
In this section, we have simply summarized the results of  \cite{qfluid,qfluid2,qfluideft} emphasizing their physical motivation.     We are now ready to use these results to calculate the entropy density and the viscosity.
\section{The entropy of a gas of sound-waves and vortices \label{secentropy}}
The entropy density $s$ is, for a free theory of massless spinless bosons, with a dispersion relation $E=cp$ is simply
\begin{equation}
\label{entropy}
s = \volfact \int_0^\Lambda p^2 dp f(c,p)
\end{equation}
where, in the Grand canonical ensemble for the Bose-Einstein distribution.  For zero chemical potential, this is
\begin{equation}
f(c,p) = \frac{cp}{T} \frac{e^{-E/T}}{1-e^{-E/T}} + \ln\left( 1-e^{-E/T} \right)
\end{equation}
The entropy of a ``thermalized gas of sound waves and vortices'', described in the previous section is then
\begin{equation}
\label{eqs}
s = \volfact \int_0^\Lambda p^2 dp \left(  f(c_L,p)+f(c_T,p) \right)
\end{equation}
and $L$ and $T$ are the transverse and longitudinal phonons.
Since higher order corrections to $s$ are irrelevant when $\eta/s$ are calculated to first order, the free theory estimate is enough.
However,  $c_T \rightarrow 0$ means Bose-Einstein corrections become important at 
 {\em arbitrary} temperature.   Hence, for vortices, the classical Maxwell-Boltzmann distribution is never a good approximation \cite{huang}.
We also introduce a cutoff $\Lambda$, which could be brought to infinity for really ideal fluids or kept large but finite if this is an EFT-type expansion, as suggested in \cite{qfluideft}.

Note that $\beta=1/T$ is {\em not} the temperature of the microscopic degrees of freedom but the temperature of a {\em gas of sound waves and vortices}.   
In an imperfect fluid, the two scales (the gas of sound-waves and vortices and the microscopic degrees of freedom) mutually thermalize after a time $\sim Ts/\eta$ (In the $g\gg 1$ limit all vortices and sound-waves just dissipate, for $g \sim 1, T\sim E_{total}$ the system converges to a thermalized fluid with finite energy, momentum and angular momentum described in \cite{becattini} ).

In an ideal fluid, 
the two systems  {\em never} reach thermal equilibrium, as can be understood from the fact that the microscopic and macroscopic scales are infinitely separated in ideal hydrodynamics.
This is why we ignore the microscopic contribution to the entropy rather than adding it to Eq. \ref{entropy}  (thereby avoiding the issue of double-counting entropy for finite $g$).   This is also why, in the next section, we shall ignore the microscopic viscosity (usually, in multicomponent systems the smallest viscosity dominates.  In our fluid, therefore, the ``zero'' microscopic viscosity should dominate over the finite viscosity carried by the sound waves).

We nevertheless equalize the microscopic and microscopic temperature as an {\em initial condition}, and therefore assume the validity of the Gibbs-Duhem relation Eq. \ref{gibbs} also for the {\em macroscopic} gas.   
If we do not, our final $\eta/s$ will depend on the arbitrary ratio of temperatures at the two scales.

\section{The viscosity of a gas of sound waves and vortices}
Comparing Eq. \ref{ttcorr} with the Kubo formula  \cite{kubo1,kubo2,kubo3}
\begin{equation}
\label{etakubo}
 \eta
 =
 {\beta\over 20}
 	     \lim_{\omega \to 0}
 		\lim_{ {\bf q} \to 0 }
\int d^3 {\bf x}\, dt\, e^{-i{\bf q}{\cdot}{\bf x} +i\omega
 t}\,
     \ave{ \pi_{lm}(t,{\bf x}) \pi_{lm}(0,{\bf 0}) }_{\rm eq}
\end{equation}
where $\ave{X}_{eq}$ refers to thermodynamic equilibrium and
where $\pi_{lm}$ are the appropriate components of the energy-momentum tensor $T_{\mu \nu}$
\[\
\pi_{lm}(x) \equiv T_{lm}(x)
 - {\textstyle{1\over 3}} \delta_{lm} T_i^i(x)
 \;
\]
makes it immediately apparent that quantum corrections to an ``ideal'' fluid will inevitably give rise to viscous terms.
  This is a ``quantum'' restatement of the result quoted in \cite{mooresound0,mooresound}, that when viscosity is small enough a significant contribution to both entropy and viscosity will be carried by {\em sound waves}.  Eq. \ref{etakubo}  will therefore need to be renormalized to take the effect of backreaction by sound-waves and vortices into account.   

To do this, one needs to compute Eqs \ref{etakubo}  not for a gas of microscopic constituents, but for a gas of sound waves and vortices, using the Lagrangian for a quantized fluid Eq. \ref{lfluid} or it's ``Jelly'' extension Eq. \ref{lfluidct}.

\begin{figure*}[t]
\begin{center}
\hspace*{2cm}
\psfig{width=17cm,figure=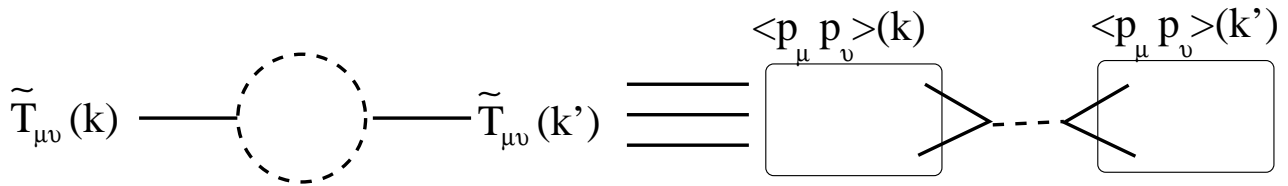                 }
\caption{\label{figkubo}
Diagrammatic illustration for the equivalence between Kubo and Boltzmann relations for the transport coefficient at tree level.   See \cite{jeon} for a derivation
}
\end{center}
\end{figure*}

  Calculating the renormalized $\eta$  from Eq. \ref{etakubo} would be a very non-trivial exercise.    Luckily it can be shown \cite{jeon} that the Kubo formulae to leading order ( 1-loop. At tree level $T_{\mu \nu}$ in the grand canonical ensemble are uncorrelated at large distance separations) correspond to the Boltzmann equation estimate for $\eta$ \cite{huang,weinberg} (at tree level).  This can be thought of as a finite temperature analogue of the optical theorem (See Fig. \ref{figkubo})

Fortunately, given the results in \cite{qfluid}, doing this is a relatively straight-forward exercise.
Hence, we have \cite{huang,weinberg}
\begin{equation}
\label{etalmfp}
\eta = \frac{1}{3} \ave{n} \ave{p} l_{mfp} = \frac{1}{3} \ave{p} \frac{\left( \sum_i \ave{n_i}\right)^2}{\sum_{ij} \ave{n_i}\ave{n_j} \ave{\sigma}_{ij \rightarrow \forall}}  
\end{equation}
Here, $l_{mfp}$ is the mean free path and $\sigma$ the interaction cross-section of the gas of hydrodynamic degrees of freedom.

In accordance with \cite{qfluid}, we include both sound waves (with speed of sound $c_s$) and ``jelly-like'' vortices (with speed of sound $c_T$).
Hence, the shear viscosity formula becomes (note that in the grand canonical ensemble correlations between different $\ave{n_i}$s vanish)
\begin{equation}
\label{eqeta}
\eta = \frac{1}{3}  \frac{\left( \ave{p_L} + \ave{p_T}  \right) \left(\ave{n_L}+\ave{n_T}\right)}{\ave{\sigma}}
\end{equation}
\begin{equation}
\ave{\sigma} =  \dave{\sigma}_{LL \leftrightarrow LL} + \dave{\sigma}_{LL \leftrightarrow LT} \end{equation} \[\ + 2 \dave{\sigma}_{LT \leftrightarrow LT} + \dave{\sigma}_{LT \leftrightarrow LL}   +  \dave{\sigma}_{TT \leftrightarrow TT}
\]
where the thermal averaging for 2-particle processes is
\begin{equation}
\label{dave}
 \dave{\sigma_{XY \leftrightarrow AB}} = \int d^3 p_1 \int d^3 p_2 n(p_1) n(p_2) \sigma_{12 \rightarrow AB}(\vec{p_1} - \vec{p_2})  
\end{equation}
Note the $\leftrightarrow$, since ideal hydrodynamics is time-reversible.
\begin{figure*}[t]
\begin{center}
\hspace*{2cm}
\psfig{width=17cm,figure=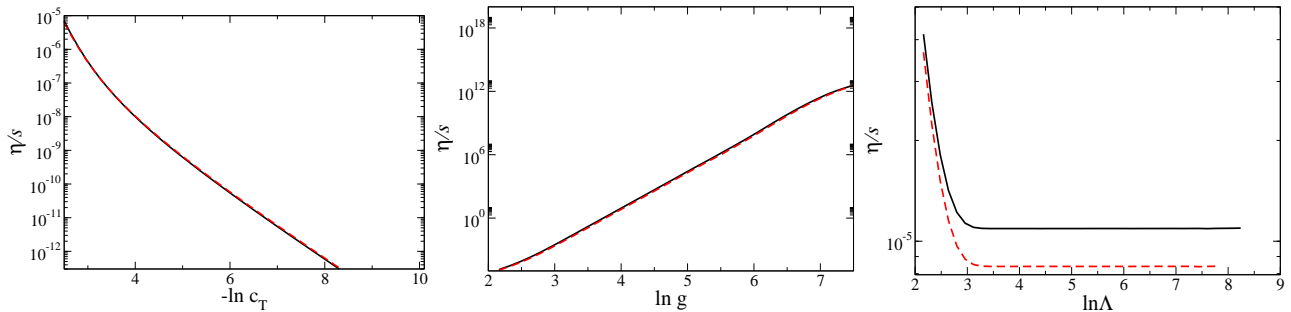                 }
\caption{\label{figetabare}
(color online) $\eta/s$, in arbitrary units, as a function of $c_T,g,\Lambda$ for the two equations of state (ideal EoS in a continuous black line, cross-over EoS for a dashed red line).  The other parameters are fixed at unity in arbitrary units.  The plot is truncated when, at large values of $g$ and high values of $\Lambda$, the calculation of $\eta/s$ becomes numerically unstable because the integrand from 0 to $\Lambda$ is dominated by very small values}
\end{center}
\end{figure*}
For the qualitative estimate, the decay processes examined in \cite{qfluid} (e.g. $L \rightarrow LT$) play no role, but they will renormalize the ``mass'' and ``coupling constant'' of the soundwave in a way that, in a finite temperature environment, needs to be resummed \cite{jeon,arnold,amy,danielf}.  These corrections, $\sim \ln c_{s},\ln c_T, \ln f_n$ are left for further work.

The bulk viscosity $\zeta$ is related to the shear viscosity by the classical formula \cite{weinberg,lifs}
\begin{equation}
\zeta = \frac{1}{3} \eta \left( 1 -  c_{ss}^2  \right)^2
\end{equation}
where $c_{ss}$ is {\em not} the microscopic speed of sound, but rather the speed of a ``sound wave in a gas of sound-waves an vortices''.    For a gas of sound-waves of speed $c_s$ this will not be equal to $1/\sqrt{3}$ since the dispersion relation of the sound waves is not $E=p$, but will have to be calculated by a formula such as
\begin{equation}
c_{ss}^2 = \frac{d \ln T_{s}}{d \ln s_{s}}
\end{equation}
where again $T_{s}$ and $s_{s}$ are the temperature and entropy of the gas of sound waves.
Thus, quantizing hydrodynamics inevitably breaks any conformal symmetry present at the classical level.

The interaction cross-section is related to the matrix elements calculable in quantum field theory using the usual relations, and has been calculated in \cite{qfluid} as
\begin{equation}
\sigma_{AA\rightarrow BB} = \alpha_{AA \rightarrow BB} \Phi (p,w_0)
\end{equation}
where the kernel is
\begin{equation}
\label{fp}
\Phi\left( p,w_0 \right) = \frac{1}{p^2}\left( \frac{p^4}{w_0 } \right)^2
\end{equation}
and the prefactors are
\begin{equation}
\label{alphatt}
\alpha_{TT\leftrightarrow TT}=\frac{1}{256 \pi} \left( \frac{13}{15}\right) \frac{1}{c_T^2} 
\end{equation}
\begin{equation}
\alpha_{LT \leftrightarrow LT}= \frac{1}{105\pi} \frac{1+7c_s^4}{c_s^2} +\mathcal{O}(c_T) 
\end{equation}
\begin{equation}
\alpha_{LL \leftrightarrow LT} \sim  \frac{c_T}{c_s^3} 
\end{equation}
\begin{equation} \label{sigmaL}
\alpha_{LL\leftrightarrow LL}=\frac{1}{256 \pi c_s^2}\left[2 \alpha^2+\frac{4 \alpha \beta}{3}+\frac{2 \beta^2}{5}\right]  
\end{equation}
where $p$ is the exchanged momentum, $w_0$ is the {\em microscopic} enthalpy density ($=Ts$) of the background fluid and 
\begin{equation}
\alpha=  f_4/c_s^2  -2f_3^2/c_s^4  + 3 c_s^2 + 2 f_3 + c_s^4
\end{equation}
\begin{equation} 
\beta = 2(1-3c_s^2)
\end{equation}
Note the $\sim c_{T}^{-2}$ divergence of $\alpha_{TT \leftrightarrow TT}$, the cross-section of the scattering of two sound waves by the production of an intermediate vortex.   Physically, this is a manifestation of the fact that such vortices can, in the hydrodynamic limit, live for an arbitrarily long time, spoiling the ``free sound-wave'' approximation.     For finite $g$, this could be an indication the quantum hydrodynamic ground-state is non-trivial.  We will see, however, that this divergence can be cancelled by a divergence in $g$.
\section{Results and discussion}
We consider two equations of state:  One ideal, one which interpolates the QCD cross-over \cite{choj,huov} and is qualitatively equivalent to recent lattice QCD studies \cite{pollat1,pollat2}.
\begin{equation}
\label{eoseq}
F(B) = \left\{  \begin{array}{c}
B^{2/3}\\
f_1 B^{\zeta}/B_0^{f_2/2-2/3} 
\end{array}  \right.
\end{equation}
where the fitted parameters are  $B_0 = \Lambda_{QCD}^6,f_1 \simeq 1.16,f_2 \simeq 0.85$.

A plot for $\eta/s$ with $B=1$ and varying $c_T^{-1},g,\Lambda$ is shown in Fig. \ref{figetabare} for both equations of state in Eq. \ref{eoseq}.   It is apparent that, unsurprisingly given the formulae of the previous section $\eta/s \rightarrow 0$ whenever $c_T^{-1}$ and $\rightarrow \infty$ when $g \rightarrow \infty$.
$\Lambda$, however, can go to $\infty$ while $\eta/s$ maintains a finite value, albeit smaller by orders of magnitude than previous lower limits on viscosity \cite{daniel,4pi}.
\begin{figure}[h]
\psfig{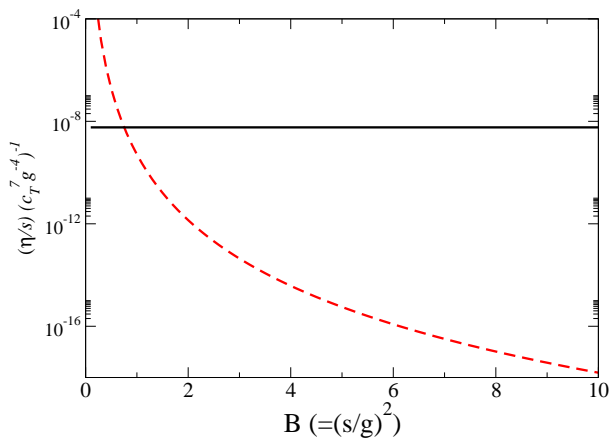}
\caption{\label{figfinal}
(color online) ``renormalized'' $\eta/s$ the two equations of state (ideal EoS in a continuous black line, cross-over EoS for a dashed red line)  }
\end{figure}

As argued in \cite{mooresound}, the divergence of $g$ is necessary to eliminate {\em statistical} hydrodynamic fluctuations, something that is needed to make the quantization procedure described here meaningful, as for any finite $g,T$ volume elements would contain only finite numbers of degrees of freedom.   Of course, a divergence in $c_T^{-1}$ is needed to reach the fluid limit from a jelly.

At the moment, we do not have a convincing argument as to why divergences in $\Lambda,g,c_T$ should be correlated.   However,
inspired by the renormalization procedure, one can ask whether one can make $c_T^{-1},g,\Lambda$ diverge so as to cancel the divergences in $\eta/s$, which would then reach a ``fixed-point'' value.
Note that, while this looks like renormalization, it is different in that, as pointed out in \cite{qfluid}, a finite $c_T$ modifies the {\em infrared} behavior of the theory while keeping the ultraviolet limit invariant.  In the conclusion we will discuss on how to go beyond this procedure, but, as we will show, this procedure does provide a recipe of eliminating $c_T$ and $g$ dependence from $\eta/s$, at least to one-loop order.

To proceed, we note that at $\Lambda/T \rightarrow \infty$ the integrals shown in the previous two sections become analytically solvable.   The key ingredient is that the thermal average (defined in Eq. \ref{dave}) of Eq. \ref{fp} is given by
\begin{equation}
\dave{\Phi(p,w_0)}_{c_1,c_2}= \volfactsq 80640 H(c_1,c_2) \frac{T^{12}}{w_0^2}
\end{equation}
\[\ H(c_1,c_2)=\frac{  \left(2 \zeta (3) \zeta (9)
   \left(c_1^6+c_2^6\right)+3 c_1^2
   c_2^2 \zeta (5) \zeta (7)
   \left(c_1^2+c_2^2\right)\right)}{  c_1^9
   c_2^9}\]
where $c_{1,2}$ can be $c_{T,s}$ depending on which $\sigma$ is being thermally averaged.

Similarly, $\ave{n}$,$\ave{p}$ and $s$ can be calculated in closed form for propagation speeds $c$:
\begin{equation}
\ave{n}_{c} = \volfact \frac{2}{c^2} \zeta(3) T^3
\end{equation}
\begin{equation}
\ave{p}_{c} = \volfact \frac{\pi^4}{15 c^4} T^4
\end{equation}
\begin{equation}
s_c = \volfact \frac{4 \pi^4}{45 c^3} T^3
\end{equation}
Putting together the formulae above with Eq. \ref{eqs} and Eq. \ref{eqeta} we have
\begin{equation}
\frac{\eta}{s} = \frac{\left({\frac{\order{1}}{c_s^4}+\frac{\order{1}}{c_T^4}}\right)\left({\frac{\order{1}}{c_s^3}+\frac{\order{1}}{c_T^3}}\right)}
{\frac{\order{1}}{c_s^3}+\frac{\order{1}}{c_T^3}} \frac{w_0^2}{T^8}\times
\end{equation}
\[\  \times  \left(  \alpha_{LL \rightarrow LL} H(c_s,c_s)+\alpha_{TT \rightarrow TT} H(c_T,c_T) + \right. \] 
\[\ \left. + \alpha_{LT \rightarrow LT} H(c_T,c_s)+ \alpha_{LL \rightarrow LT} H(c_s,c_s) \right)^{-1}  \]
Unless the equation of state exhibits a first-order phase transition, $c_s$ remains finite throughout, and the only divergences exhibited in this expression are in $c_T$ and $g$.    In this regime, the divergences go as
\begin{equation}
\frac{\eta}{s} \sim \frac{\order{w_0^2}}{c_T^4 T^8}\left(\underbrace{ \order{1}}_{LL \rightarrow LL}+\underbrace{\frac{\order{1}}{c_T^{14}}}_{TT \rightarrow TT} +\underbrace{\frac{\order{1}}{c_T^{9}}}_{LT \rightarrow LT} + \underbrace{\order{c_T}}_{LL \rightarrow  LT}  \right)^{-1}
\end{equation}
Remembering that 
\[\ T = \frac{w_0}{s} \sim \frac{\sqrt{B} (dF/dB)}{g}   \]
we have the final result of this work, the dependence of $\eta/s$ on $g,c_T$ and the equation of state
\begin{equation}
\label{final}
\frac{\eta}{s} = K_0 \frac{c_T^{14} g^8}{B^2 (dF/dB)^6}
\end{equation}
Hence, $c_T^7 g^4 \sim \order{1}$ ensures convergence of $\eta/s$.
In this case we lose predictivity of $\eta/s$ at a given entropy $\sqrt{B}$.   The {\em variation} of $\eta/s$ as a function of $\sqrt{B}$ is however only a function of the equation of state (analogously to the $\beta$-function), and hence can be unambiguously computed.      The numerical constant $K_0$, counting all factors of $\pi,\zeta,...$ is 
\[\ K_0= \frac{ \zeta(3)^2 \zeta(9)}{80640}\frac{4}{256 \pi}\frac{13}{45} \frac{\pi^2}{15} \left( \frac{4\pi^4}{45} \right)^{-1} \simeq 1.96(10^{-9}) \]
In Fig. \ref{figfinal} we compute it for the two equations of state shown in  Eq. \ref{eoseq}. As expected, the conformal equation of state also gives a constant $\eta/s$.   We see that this limit is well below those considered previously in the literature \cite{daniel,4pi,beyond4pi} but is nevertheless finite.

Counting the microscopic entropy 
destroys any non-trivial limit of this calculation, as we have {\em two} potential divergences:
\begin{equation}
\frac{\eta}{s} \sim \frac{w_0^2}{T^{5}}\frac{c_T^{11}}{c_T^{-3} T^3+s_{micro}} \sim \frac{c_T^{11}}{c_T^{-3}g^{-8} B^2 \left(\frac{dF}{dB} \right)^6+g\sqrt{B}}
\end{equation}
generally, this $\rightarrow 0$ as $g \rightarrow \infty$ or as $c_T \rightarrow 0$ no matter how the other quantity behaves.   In this limit, therefore, one recovers something like the ``many-species fluid'' of \cite{cohen1,cohen2}.
As argued in section \ref{secentropy}, however, in the ideal fluid limit, when the time required for microscopic and macroscopic degrees of freedom to ''talk'' diverges, counting the microscopic entropy is not well-motivated


The estimates conducted here are without doubt extremely rough.   They lack, for instance, the resummation-derived log terms in \cite{arnold,jeon,amy,danielf}.
They are however sufficient to draw some qualitative conclusions, for
the point of this exercise was to investigate whether "promoting hydrodynamic perturbations to quantum degrees of freedom" yields a way of extracting a "quantum bound" for $\eta/s$ insensitive to the "UV" details of the system (the microscopic theory).    At 1-loop, the answer seems to be yes. 

 Since this theory is non-renormalizeable, and since the convergence of $\eta/s$ in a perturbative series is dubious \cite{arnold,linde,htl,nansu,moore2ndorder}, it is unlikely that this result will stay invariant once higher-order corrections are added.    

Non-perturbative corrections are also likely to make an appearance, since, as argued in \cite{qfluid}, the vacuum in $c_T \rightarrow 0$ limit is inevitably dominated by strongly-interacting quantum vortices.   Since several analytical solutions of ideal fluid mechanics are known, an estimate of  non-perturbative contributions to viscosity could perhaps be obtained by constructing semi-classical ``instantons'' between solutions of different entropy content.   These will be explored in \cite{comingwork}.    A more systematic approach would be to put this theory on the lattice, in order to both see how far is the vacuum state from the ``hydrostatic one'', and to find out whether the theory has a ``trivial'' continuum limit.    
Since, as argued throughout this work,  the ratio of the microscopic to macroscopic entropy (parametrized here by $g$) is as important as the mean free path in determining quantum fluctuations, an intriguing possibility is that $g$ drives a phase transition between a "classical vacuum state" and a "quantum-turbulent" vacuum state, analogously to the finite density phase transition in the number of colors ($g\sim N_c^2$ in Gauge theory) discussed in \cite{nc1,nc2}.

The discussion in this work suggests that, as we long suspected and as results such as \cite{daniel,4pi} illustrate, {\em some} entropy generation is inevitable when the system is quantized.  Naively, this seems to be at odds with the result, shown by Von Neumann, that entropy, rigorously defined in terms of the density matrix $\hat{\rho}$ as $S=-\mathrm{Tr}\left( \hat{\rho} \ln \hat{\rho} \right)$, is conserved during the quantum evolution of a system \cite{entropy1}.   This apparent paradox has given rise to quite a lot of theoretical activity, from heavy ions to black holes and condensed-matter systems (see, e.g., \cite{entropy1,entropy2,entropy3})

One physical resolution to this paradox in the context of a quantum field theory is to note that the Hilbert space goes to arbitrary high momenta, but any conceivable experiment has a finite-momentum resolution,parametrized by a scale $\Lambda$.    Observables, therefore, must inevitably be {\em renormalized}, with a division into {\em slow} degrees of freedom (which we observe) and fast ones (at much higher energy $\Lambda$ than any scale of our detector, they can only appear as virtual states suppressed by powers of $\Lambda^n$) \cite{peskin}.

For matrix elements with definite numbers of particles (``n-point functions'') this cut-off can be absorbed into a redefinition of the Lagrangian. For observables tracing over an undefined number of particles (such as the thermally averaged quantities, $\ave{...}$ and $\dave{...}$ described in this work) the cut-off inevitably introduces dissipative terms into the observable's equation of motion.    This is obvious from the fact that Wilson's coarse-graining  \cite{peskin} looks exactly like decoherence  \cite{entropy3}, with the fast degrees of freedom playing the role of the environment and the slow ones of the system.  
These terms will also be suppressed by powers of $\Lambda$, but the example of turbulence shows that a small viscosity does not necessarily mean small entropy generation.    

 While we can not at the moment prove that $c_T^{-1},g\rightarrow \infty$ corresponds to a renormalization group flow, the fact that a non-zero $\eta/s$ arises out of a quantized ideal fluid suggests that the considerations above apply  whether the system's observable degrees of freedom are ``microscopic'' particles or ``macroscopic'' collective excitations such as sound-waves and vortices.

In conclusion, we discussed the effective $\eta/s$ of a quantum ideal fluid.   We have shown that, in general, a finite $\eta/s$ can be generated by quantum sound and vortex excitations even if at the classical level the system's equations of motion correspond to an ideal ($\eta/s=0$) fluid.    We estimated this $\eta/s$ as a function of the equation of state to first order in perturbation theory, and discussed the physical meaning of our results.

 G.T. acknowledges the financial
support received from the Helmholtz International Center for FAIR within the framework of the LOEWE program (Landesoffensive zur Entwicklung
Wissenschaftlich-\"Okonomischer Exzellenz) launched by the State of Hesse, as well as the Extreme Matter Institute (EMMI).
We thank Alberto Nicolis, Lam Hui, Solomon Endlich, Junpu Wang, Daniel Fernandez-Fraile, Dirk Rischke and Giuseppe Colucci for discussions.
The hospitality of Columbia university during the time in which part of this work was done is greatly appreciated.

\vskip 0.3cm

\end{document}